# Fundamental Limits of the Dew-Harvesting Technology

Minghao Dong,[†] Zheng Zhang,[†] Yu Shi,[‡] Xiaodong Zhao,[†] Shanhui Fan,[‡] and Zhen Chen[*, †]

[†]Jiangsu Key Laboratory for Design & Manufacture or Micro/Nano Biomedical Instruments, School of Mechanical Engineering, Southeast University, Nanjing 210096, China

[‡]Ginzton Laboratory, Department of Electrical Engineering, Stanford University, Stanford, CA94305, USA

*Supporting information*

**ABSTRACT:** Dew-harvesting technology radiatively cools a condenser below the dewpoint to achieve condensation of the water vapor from the atmosphere. Due to its passive nature, this technology has attracted a broad interest, in particular in the context of the worldwide drinking-water scarcity. However, the fundamental limit of its performance has not yet been clarified. Moreover, the existing applications have been limited to humid areas. Here, we point out the upper bound of the performance of this technology by carefully considering the spectral directional atmospheric transmittance in a wide range of parameters such as the ambient temperature ($T_{ambient}$), the relative humidity (RH), and the convection coefficient ($h$). Moreover, we highlight the potential of a condenser consisting of a selective emitter, which is capable of condensing water vapor under significantly more arid conditions as compared with the use of a blackbody emitter. For example, a near-ideal emitter could achieve a dew-harvesting mass flux ($\dot{m}''$) of 13 gm$^{-2}$hr$^{-1}$ even at $T_{ambient}$ = 20 °C with RH = 40%, whereas the black emitter cannot operate. We provide a numerical design of such a selective emitter, consisting of six layers, optimized for dew-harvesting purposes.

**KEYWORDS:** dew-harvesting, radiative cooling, selective emitter, drinking water,

Sustainable access to fresh water has been recognized as one of the grand challenges for engineering in the 21$^{st}$ century.[1] Traditional technologies, such as distillation,[2] reverse osmosis,[3] and waste water recycling,[4] are energy-consuming and expensive. It is therefore difficult to widely apply them especially in developing countries where the crisis to access fresh drinking water is the most serious. Recent passive techniques utilizing solar energy,[5-9] on the other hand, are constrained either by the exotic and expensive water absorbing material[5-6] or by the bulky solar concentrating components.[8-9]

Dew-harvesting technology utilizes the ultracold outer space to radiatively cool a surface below the dewpoint and condenses the water vapor from the atmosphere. This passive technology holds great potential for fresh water harvesting due to the fact that a significant amount of water vapor is stored in the atmosphere.[10] The theoretical analysis on dew-harvesting has a long history. Following the pioneering work by Lewis on the evaporation of a liquid into a gas,[11] Hofmann developed an analogous mathematical framework to analyze the heat and mass transfer in the process of the dew formation.[12] Since then, several theoretical works[13-19] have been conducted to quantitatively explore the dew-harvesting. However, none of these works takes the spectral atmospheric transmittance into account, and very few of them considers the spectral emissivity of the condenser.[14-15] As a result, the fundamental limits of this technique have not been properly clarified, making it hard to evaluate the performance of the experiments[13-14, 20-22] and to determine whether or not this technology is applicable under various conditions, in particular in relatively arid areas.[5-6]

In this paper, building upon the recent developments of the radiative cooling technology,[23-30] we develop a theoretical framework to analyze the dew-harvesting mass flux ($\dot{m}''$) of a black and a selective emitter in a wide range of parameters including the ambient temperature ($T_{ambient}$), the relative humidity (RH), and the convection coefficient ($h$). We point out that the selective emitter surpasses its black counterpart, in particular under high $T_{ambient}$ in relative arid areas, where the needs of the potable water are more demanding but impossible to be met using a black emitter. For example, at $T_{ambient}$ = 20 ℃ with RH = 40%, a black emitter fails in harvesting water by any means; in contrast, the selective emitter could have a $\dot{m}''$ of 13 gm$^{-2}$hr$^{-1}$. To realize the potentials of a selective emitter, we design a simple multilayer stack to demonstrate its dew-harvesting abilities. We also quantify the bifunction of the convection, and propose a setup to adjust $h$, for the purpose of maximizing $\dot{m}''$.

## ■ RESULTS

**Theoretical Framework.** Figure 1(a) shows a typical setup for dew-harvesting in most existing experiments,[17-18, 20] in which the condenser is radiatively cooled to be below the dewpoint to condense the water vapor from the adjacent air. The effects of the convection over the top surface of the condenser are twofold: it recharges the condenser with fresh water vapor, which is beneficial for dew-harvesting, but also warms the condenser, which is detrimental. Therefore, there must exist an optimized convection coefficient, $h$, that maximizes $\dot{m}''$. We will return to this insight shortly in the context of Fig. 5. As a result, we propose a second setup (Fig. 1b), which is equipped with a low-power air pump to actively control the air speed and thus $h$.

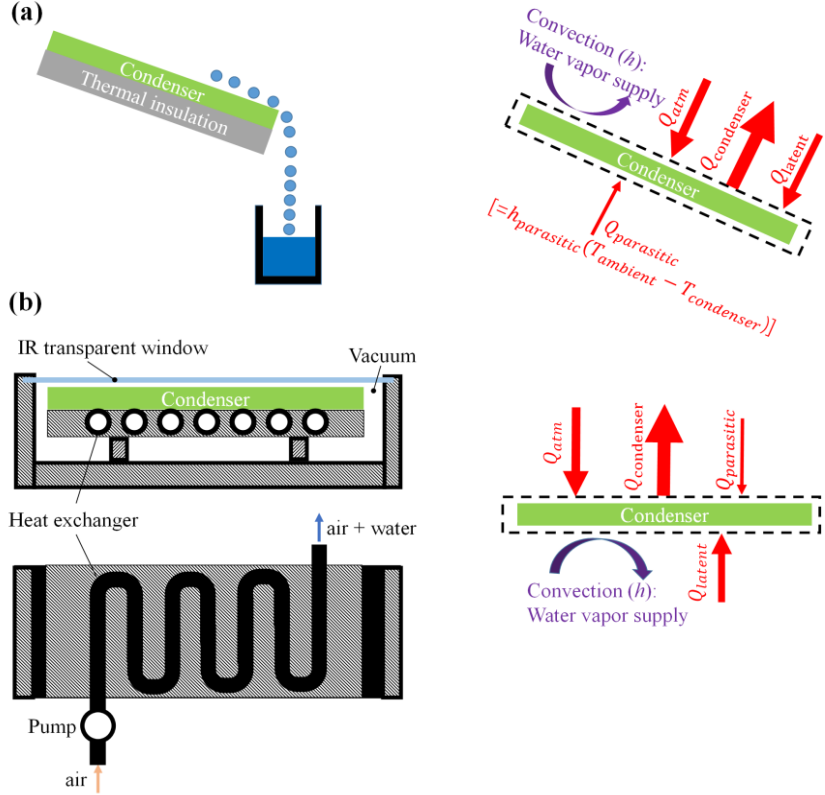

**Figure 1**. Two setups for dew-harvesting, both of which utilize the radiative cooling technology to cool the condenser below the dewpoint temperature and condense the water vapor from the air. (a) Passive setup and the corresponding energy balance (see text): top surface of the condenser is used for both cooling and water collection. While the parasitic heat loss from the bottom surface, characterized by a convection coefficient, $h_{parasitic}$, weakens the cooling performance and thus reduces water production, the convection from the top surface, characterized by a different convection coefficient, $h$, has one additional function: supplying fresh air with water vapor. (b) Active setup and the corresponding energy balance (see text): top and bottom surfaces are for cooling and water collection, respectively. In this scenario, fresh air with water vapor is pumped into a heat exchanger underneath the condenser, which also weakens the cooling performance of the condenser. The air conduction/convection from the top surface is a pure parasitic heat loss, which can be significantly reduced by evacuating the enclosure to vacuum.

To obtain both the temperature of the condenser, $T_{condenser}$, and the mass flux of dew-harvesting, $\dot{m}''$, we require two constraints. The first one comes from the energy balance (right panels of Fig. 1) of the condenser, in which the latent heat released by the condensation process can be expressed as

$$Q_{latent} = Q_{condenser}(T_{condenser}) - Q_{atm}(T_{ambient}, \text{RH}) - (h + h_{parasitic})(T_{ambient} - T_{condenser}), (1)$$

where $Q_{condenser}$ and $Q_{atm}$ represent the thermal radiation emitted from the condenser and absorbed from the atmosphere, respectively.[23, 25] The parasitic heat transfer coefficient, $h_{parasitic}$, accounts for all the parasitic heat transfer between the condenser and the environment except for the convection that carries water vapor to the condenser, which is characterized by $h$. For theoretical upper bounds, we set $h_{parasitic} = 0$ in all calculations in the main text, leaving the discussion on the effect of a nonzero $h_{parasitic}$ in Supporting Information Figure S3 and Note 2. Note here we choose the parameter pair

($T_{ambient}$, RH), where RH is the relative humidity, instead of ($T_{ambient}$, PWV), where PWV stands for the precipitable water vapor, to quantify the atmospheric transmittance and thus $Q_{atm}$,[31] because the relative humidity is the most straightforward indicator for dew-harvesting, as will be evident in the following discussion.

For the second constraint, we follow the analyses of Lewis[11] and Hofmann.[12] Analyzing the heat and mass transfer simultaneously,[32] we develop a general expression

$$Q_{latent} = (Le)^n \frac{h}{\gamma}[\text{RH} \times P_{H_2O}(T_{ambient}) - P_{H_2O}(T_{condenser})], \quad (2)$$

which is valid in various situations characterized by different power, $n$, of the Lewis number, $Le$ (= 0.87 at 300 K),[32] as discussed in detail in METHODS and Supporting Information Note 2. Corresponding to the setup in Fig. 1(a), we set $n$ = -2/3 and thus $(Le)^n$ = 1.1 for the calculations in this work, which introduces an error less than 6% if we consider the other scenarios, e.g. the setup in Fig. 1(b). Here $\gamma$ is the psychrometer constant (67 PaK$^{-1}$ at 20°C),[33] and $P_{H_2O}$ is the saturation vapor pressure, which depends only on the temperature.[34] The last term of Eq. (2) assumes that the air close to the condenser surface is saturated in water vapor, which is justified since $T_{condenser}$ needs to be lower than the dewpoint temperature, $T_{dew-point}$, for condensation to start.

Solving Eqs. (1) and (2), we obtain $T_{condenser}$ and $Q_{latent}$. The latter can be converted to the mass flux of dew-harvesting,

$$\dot{m}'' = \frac{Q_{latent}}{\Delta}, \quad (3)$$

in which $\Delta$ is the latent heat per unit mass. Figure 2 shows the RH dependence of $T_{condenser}$ (black solid line; left axis) and $\dot{m}''$ (orange solid line; right axis) of a black emitter at $T_{ambient}$ = 20°C, under a typical rooftop scenario ($h$ = 8 Wm$^{-2}$K$^{-1}$). The dewpoint temperature, $T_{dew-point}$ (black dashed line; left axis), is also calculated[35] for reference. At low RH, $T_{condenser}$ is higher than $T_{dew-point}$, and thus no dew forms. As RH increases, $T_{dew-point}$ increases faster than $T_{condenser}$, resulting in a crossover at RH = 74%, where water vapor starts to condense to liquid. Beyond this crossover, $\dot{m}''$ monotonically increases, and finally reaches 42 gm$^{-2}$hr$^{-1}$ at RH = 100%.

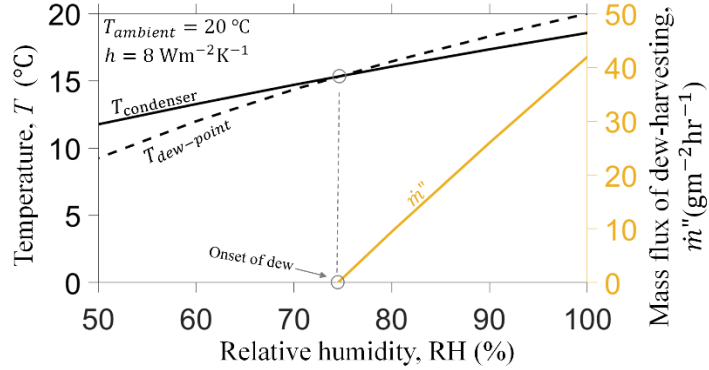

**Figure 2**. Simultaneously solving the temperature, $T_{condenser}$ (black solid line; left axis), and the condensation mass flux, $\dot{m}''$ (orange; right axis), as a function of the relative humidity, RH, for a black emitter at $T_{ambient}$ = 20 °C, under a typical rooftop scenario ($h$ = 8 Wm$^{-2}$K$^{-1}$). The dewpoint temperature (black dashed line) is also calculated for reference. Dew does not form until a characteristic relative humidity, RH=74%, at which $T_{condenser}$ starts to be lower than $T_{dew-point}$. As RH increases, $\dot{m}''$ monotonically increases to 42 gm$^{-2}$hr$^{-1}$ at RH = 100%.

**Black vs. Selective Emitter.** Figure 3 analyzes $\dot{m}''$ for three different emitters: a black emitter (black lines), a near-ideal emitter (blue lines) which emits 100% in the wavelength range of 8–13 μm and 0% elsewhere,[25] and a photonic design in this work (red lines), under two representative scenarios: a typical rooftop setup ($h$ = 8 Wm$^{-2}$K$^{-1}$), e.g. in Fig. 1(a), and a setup that reduces the wind speed ($h$ = 2 Wm$^{-2}$K$^{-1}$), e.g. in Fig. 1(b). We will return to the photonic design shortly in the context of Fig. 4. The dominant feature of Fig. 3 is that the near-ideal emitter surpasses its black counterpart in any circumstance. In particular, the near-ideal emitter is capable of harvesting dew at much lower RH (Fig. 3a) and much higher $T_{ambient}$ (Fig. 3b). Table 1 presents two examples of this feature. First, in a relatively dry condition (RH = 40% and $T_{ambient}$ = 20 °C; first column), the black emitter does not work at all, while the near-ideal one has a $\dot{m}''$ of 13 gm$^{-2}$hr$^{-1}$. Second, in a high temperature condition ($T_{ambient}$ = 30 °C and RH = 60%; second column), the black emitter again fails in harvesting water, while the near-ideal one has a $\dot{m}''$ up to 14 gm$^{-2}$hr$^{-1}$. We note the comparison between the two groups of scenarios in

Fig. 3(a): $h = 8$ Wm$^{-2}$K$^{-1}$ (dashed lines) vs. $h = 2$ Wm$^{-2}$K$^{-1}$ (solid lines). While the former shows less difference among the three emitters, the latter amplifies the advantages of the selective emitters, in particular at low RH. Dashed lines in Fig. 3 represent $\dot{m}''$ values that would have been reached if dew formed. However, under these conditions, $T_{condenser}$ is below 0℃, and thus frost forms, which hinders the water-harvesting process, in particular in the setup of Fig. 1(a).

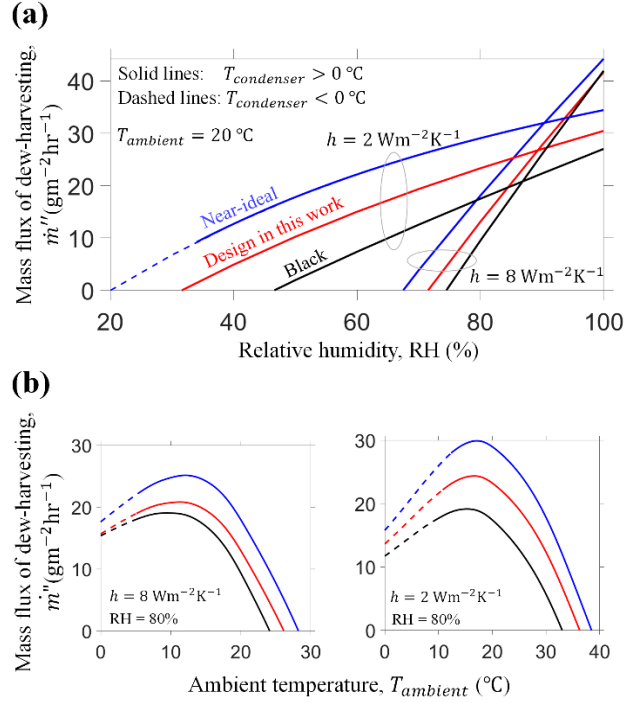

**Figure 3**. Relative humidity (RH) and ambient temperature ($T_{ambient}$) dependence of the mass flux of dew-harvesting ($\dot{m}''$). (a) RH dependence of $\dot{m}''$ for three different emitters at $T_{ambient}$ = 20 °C under two representative scenarios: $h = 2$ Wm$^{-2}$K$^{-1}$ and 8 Wm$^{-2}$K$^{-1}$. For any emitter under any scenario, $\dot{m}''$ increases monotonically as the increase of RH. The near-ideal emitter (blue) surpasses its black counterpart (black) in both scenarios. In particular, the former is capable of harvesting water at significantly low RH. (b) $T_{ambient}$ dependence of $\dot{m}''$. For a fixed RH, the content of the water vapor in the atmosphere increases with the increase of $T_{ambient}$, which, on one hand, is beneficial to dew-harvesting, but, on the other hand, is detrimental due to the blurred atmospheric transparency, resulting in the reduction of the radiative cooling flux. This competition results in maximal $\dot{m}''$ at optimized $T_{ambient}$. Dashed lines represent $\dot{m}''$ values that would have been reached if dew formed, although in reality frost forms instead, because $T_{condenser}$ < 0 °C.

The other major feature is the trend that $\dot{m}''$ increases monotonically with the increase of RH in Fig. 3(a), but has maxima at optimized $T_{ambient}$ in Fig. 3(b), regardless of the type of the condenser and the value of the convection coefficient. These trends result from the variation of the water vapor in the atmosphere, whose effects are twofold. On one hand, increasing RH ($T_{ambient}$) while fixing $T_{ambient}$ (RH), the content of the water vapor in the atmosphere increases, which is beneficial to dew-harvesting. On the other hand, it could also be detrimental since the atmospheric transmission in the transparency window is reduced[31] and thus reducing the cooling flux. The former always surpasses the latter at any RH, resulting in a monotonic trend in Fig. 3(a), while these two effects are balanced at specific $T_{ambient}$, resulting in maxima at optimized $T_{ambient}$ in Fig. 3(b).

**Table 1.** Two representative examples (first and second columns), showing the advantages of the selective emitter over the black one, as well as the fundamental limit (third column) of $\dot{m}''$. While the black emitter is slightly better at ultrahigh RH and $h$ (third column), the selective emitters are significantly better in low RH (first column) and high $T_{ambient}$ (second column).

| $\dot{m}''$ [gm$^{-2}$hr$^{-1}$] | ($T_{ambient}$, RH, $h$) [℃, %, Wm$^{-2}$K$^{-1}$] | | |
|---|---|---|---|
| | (20, 40, 2) | (30, 60, 2) | (12, 100, ∞) |
| Black | 0 | 0 | 59 |
| Near-ideal | 13 | 14 | 56 |
| Design in this work (see Fig. 4) | 5 | 5 | 55 |

Figure 4(a) shows the photonic design of a selective emitter aiming to maximize the condensation mass flux, $\dot{m}''$. The memetic algorithm[36] is used to optimize the spectrum of a multilayer device that approaches the emissivity spectrum of the near-ideal emitter. This design consists of six layers, which are made of three common dielectric materials: $MgF_2$, $SiN_4$, and $SiC$, with an aluminum (Al) substrate as the back mirror. The emissivity spectrum of this design is shown in Fig. 4(b) (red), with a representative atmospheric transmittance (grey) in a relatively dry climate ($T_{ambient}$ = 10 °C and RH = 30%) for reference.[37] We note that there exists broad and large emissivity inside the atmospheric transparency window (8-13 μm), which ensures that the condensation mass flux, $\dot{m}''$, of this photonic design significantly surpasses that of the black emitter, which is evident from the comparison in Figs. 3 and 5, as well as in Table 1 (first two columns). Moreover, this design is also better than our previous designs for radiative cooling purposes,[24-25] which are targeted at achieving low temperatures (Fig. S4 of Supporting Information). Note that in Fig. 4(b) we show the emissivity and transmittance only along the zenith angle for clarity, but in the calculations of $\dot{m}''$ we use the angle-dependent properties.

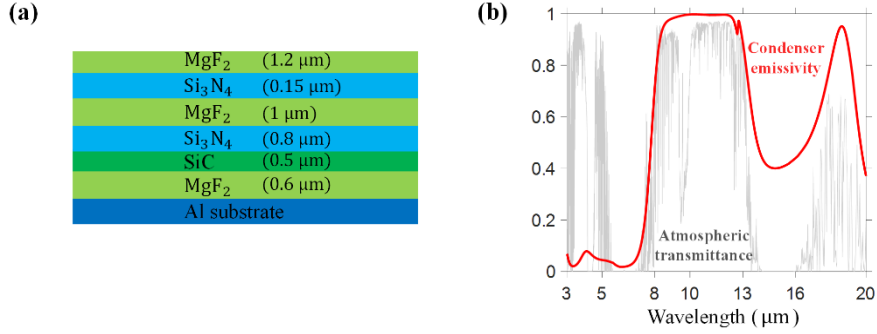

**Figure 4**. A photonic design to maximize $\dot{m}''$. (a) Schematic cross-sectional view of the multilayer stack made of three different materials ($MgF_2$, $Si_3N_4$, and SiC) on an aluminum substrate. (b) Calculated spectral emissivity of the photonic design (red) along zero zenith angle, with a typical atmospheric transmittance (grey) under a relative dry condition ($T_{ambient}$ = 10 °C and RH = 30%) for reference. Our designed structure emits significantly in the atmospheric transparency window (8-13 μm).

**Effect of Convection.** We now turn to quantifying our intuition on the effect of convection, which motivates us to propose the setup in Fig. 1(b). As an example, Figs. 5(a-b) clearly show maximal $\dot{m}''$ at optimized $h$, at $T_{ambient}$ = 20 °C with RH = 40% and 80%, regardless of the type of the condenser. These maxima originate from the two competing effects associated with convection: on one hand, convection brings water vapor to the condenser, which helps the dew-harvesting; on the other hand, convection also carries parasitic heat to reduce the cooling flux of the condenser, which hurts the dew-harvesting. As a result, at RH < 100%, $\dot{m}''$ first increases and then decreases with the increase of $h$. We note that by means of varying $h$ we not only maximize $\dot{m}''$ but also avoid frost formation (indicated by the dashed lines), which is a significant advantage of the setup we proposed in Fig. 1(b).

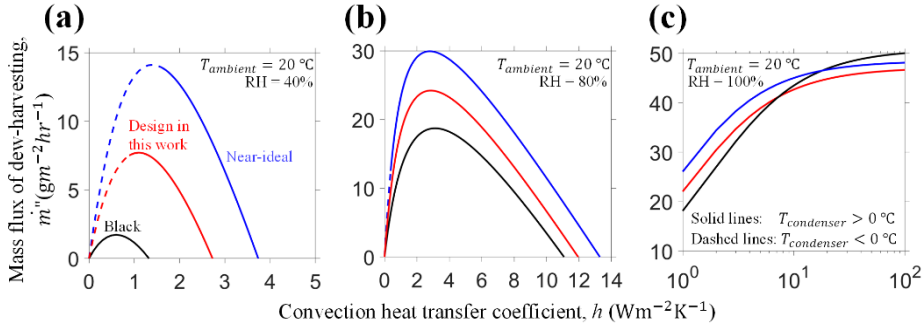

**Figure 5.** Effect of convection. (a-b) RH < 100%. As $h$ increases from zero, more and more water vapor is carried to the condenser, which increases $\dot{m}''$; as $h$ keeps increasing, however, the condenser temperature finally surpasses the dewpoint temperature ($T_{condenser} > T_{dew-point}$), which stops the condensation. (c) RH = 100%. In this case, $T_{condenser}$ can never surpass $T_{dew-point}$, since now $T_{dew-point}$ = $T_{ambient}$. Now the only role of convection is to supply water vapor, and the maximal $\dot{m}''$ is defined by the radiative cooling power at $T_{condenser}$ = $T_{ambient}$. While the selective emitters can reach significantly low temperature, the black emitter has the maximal cooling flux at $T_{condenser}$ = $T_{ambient}$, which explains the crossovers. Dashed lines represent $\dot{m}''$ values that would have been reached if dew formed. However, under these conditions, $T_{condenser}$ is below 0 °C, and thus frost forms, which hinders the dew-harvesting.

At RH = 100%, however, $\dot{m}''$ monotonically increases and finally approaches its fundamental limit at the specific $T_{ambient}$ (here, 20 ºC), as shown in Fig. 5(c). We note the logarithmic *x*-axis to emphasize the transition between the two regimes. The saturation behavior results from two facts. First, $T_{dew\text{-}point}$ strictly equals $T_{ambient}$ at RH = 100%; and second, $T_{condenser}$ approaches $T_{ambient}$ at large $h$. These two facts result in $T_{condenser} \approx T_{ambient} = T_{dew\text{-}point}$ at RH = 100% with large $h$, which requires almost no further cooling for the condensation to start. At this limit, all the radiative cooling flux is used for condensation, and thus the fundamental limit is approached at the specific $T_{ambient}$.

Another feature of Fig. 5(b) is the crossovers between the black and the selective emitters. Selective emitters have higher $\dot{m}''$ at small $h$, while black emitter is better at large $h$. This feature results from the fact that selective emitter reaches significantly lower temperature at relatively small $h$,[25] but black emitter has the maximal cooling flux at $T_{condenser} = T_{ambient}$. This result underlines that while the black emitter can reach the ultimate fundamental limit of dew-harvesting at RH = 100% and ultrahigh $h$ (3rd column of Table 1), the selective emitter is better in most other conditions (Fig. S1 of the Supporting Information).

## ■ DISCUSSION AND SUMMARY

We develop a general framework to analyze the dew-harvesting technology. In particular, we highlight the effect of important factors, such as $T_{ambient}$ and RH. Distinct from the previous analyses in the literature, we consider the wavelength and angle dependence of atmospheric transmittance. We point out that while the black emitter gives the ultimate upper bound of approximately 60 gm$^{-2}$hr$^{-1}$ at 100% of RH, a selective emitter can harvest water at much higher $T_{ambient}$ (e.g. 30 °C) with much lower RH (e.g. 40%). We design a simple multilayer stack to realize this potential. We also quantify the effect of convection, which carries not only water vapor but also parasitic heat to the emitter. Correspondingly, we propose a setup which can adjust $h$ to maximize $\dot{m}''$ and avoid frost.

We end by commenting on the prospect of this technology. Dew-harvesting could be beneficial to both humid and dry areas. The former includes islands and coastal cities which are surrounded by seawater that is not potable, while the latter include deserts which lack of any form of drinking water. Even in the hot and arid desert, there is still some amount of water vapor stored in the atmosphere, e.g. $T_{ambient}$ = 20°C and RH = 40% at night in the Mojave Desert in California,[38] which is ready to be harvested by a carefully designed photonic emitter. This passive fresh water harvesting technology would complement existing technologies, especially in rural and low-income areas where the cost is a big concern. In particularly, it could provide an option for local fresh water harvesting, which is more effective for personal needs.[1]

## ■ METHODS

We present the derivation of Eq. (2) in detail. Driven by the difference of the water vapor concentration, the mass transfer flux of the water vapor from the ambient to the condenser surface is[32]

$$\dot{m}'' = g_{m,H_2O}\left(\widehat{m}_{H_2O,ambient} - \widehat{m}_{H_2O,condenser}\right), \qquad (4)$$

where $g_{m,H_2O}$ is the mass transfer coefficient of the water vapor, with the same unit of $\dot{m}''$, and $\widehat{m}_{H_2O,ambient}$ and $\widehat{m}_{H_2O,condenser}$ are the mass fraction of water vapor in the ambient and near the condenser surface, respectively.

From the Ideal Gas Law, we write the mass fraction of water vapor as

$$\widehat{m}_{H_2O} = \frac{RH \times P_{H_2O} M_{H_2O}}{P_\infty M_{air}}, \qquad (5)$$

Where $P_\infty$ is the atmospheric pressure, $M_{H_2O}$ and $M_{air}$ are the molecular mass of water and air, respectively, and the saturation vapor pressure, $P_{H_2O}$, with unit of [$10^{-3}$bar], has an expression[34] of $P_{H_2O} = T^{a_1} 10^{c_1 + b_1/T}$, where $T$ is the absolute temperature, and the parameters are: $a_1 = -4.9283$, $b_1 = -2937.4$ K, and $c_1 = 23.5518$, which are applicable in the temperature range of -50 °C - 100 °C.

The most important step is to bridge the mass transfer coefficient, $g_{m,H_2O}$, and the convection heat transfer coefficient, $h$. Applying the strong analogy between the heat and mass transfer,[32] we obtain

$$g_{m,H_2O} = \frac{h}{c_p}(Le)^n, \qquad (6)$$

where $c_p$ is the specific heat capacity of the air at constant pressure, with unit of [Jkg$^{-1}$K$^{-1}$], and *Le* is the Lewis number, which is a measure of the relative thermal and concentration boundary layer thicknesses.[32, 39] We summarize the *n* values in Table 2 for different scenarios, including natural and forced convection at horizontal/inclined plates or in pipes with different shapes of cross section, as discussed in detail in Supporting Information Note 1. Combining Eqs. (3)-(6), we arrive at Eq. (2) of the main text.

Table 2. Values of *n* in Eqs. (2) and (6), corresponding to different scenarios. Detailed derivations see Supporting Information Note 1.

|   | Natural convection (horizontal or inclined plate) | Forced external convection (horizontal or inclined plate) | Forced internal convection (pipe) |
| --- | --- | --- | --- |
| *n* | $-3/4$ | $-2/3$ | $-1$ |

■ **ASSOCIATED CONTENT**

**Supporting Information**

Figure on the contour plots of the condensation mass flux as a function of the relative humidity (*x*-axis) and the ambient temperature (*y*-axis) under two representative scenarios.

Figure on the comparison of the condensation mass flux between the two setups in Fig. 1(a) and 1(b).

Figure on the effect of the parasitic heat transfer coefficient on the condensation mass flux.

Figure on the comparison between the photonic design in this work and the designs in literature.

Note on the relation between the heat and mass transfer coefficients.

Note on the effect of the parasitic heat transfer coefficient.

Note on the comparison between the photonic design in this work and the designs in literature.

■ **AUTHOR INFORMATION**

**Corresponding Author**
*E-mail: zhenchen@seu.edu.cn.

**Notes**
The authors declare no competing financial interest. During the preparation of this manuscript, we became aware of a related analysis.[40]

■ **ACKNOWLEDGEMENTS**

This work was supported in part by National Natural Science Foundation of China (51776038) and the Innovative and Entrepreneurial Talent Plan (Jiangsu Province, China). S. F. acknowledges the support of the U. S. Department of Energy Office of Basic Energy Sciences (grant No. DE-FG02-07ER46426).

# Fundamental Limit of the Dew-Harvesting Technology

# Supporting Information

Minghao Dong,[1] Zheng Zhang,[1] Yu Shi,[2] Xiaodong Zhao,[1] Shanhui Fan,[2] and Zhen Chen[1, *]

[1]Jiangsu Key Laboratory for Design & Manufacture or Micro/Nano Biomedical Instruments, School of Mechanical Engineering, Southeast University, Nanjing 210096, China
[2]Ginzton Laboratory, Department of Electrical Engineering, Stanford University, Stanford, CA94305, USA

*Email: zhenchen@seu.edu.cn

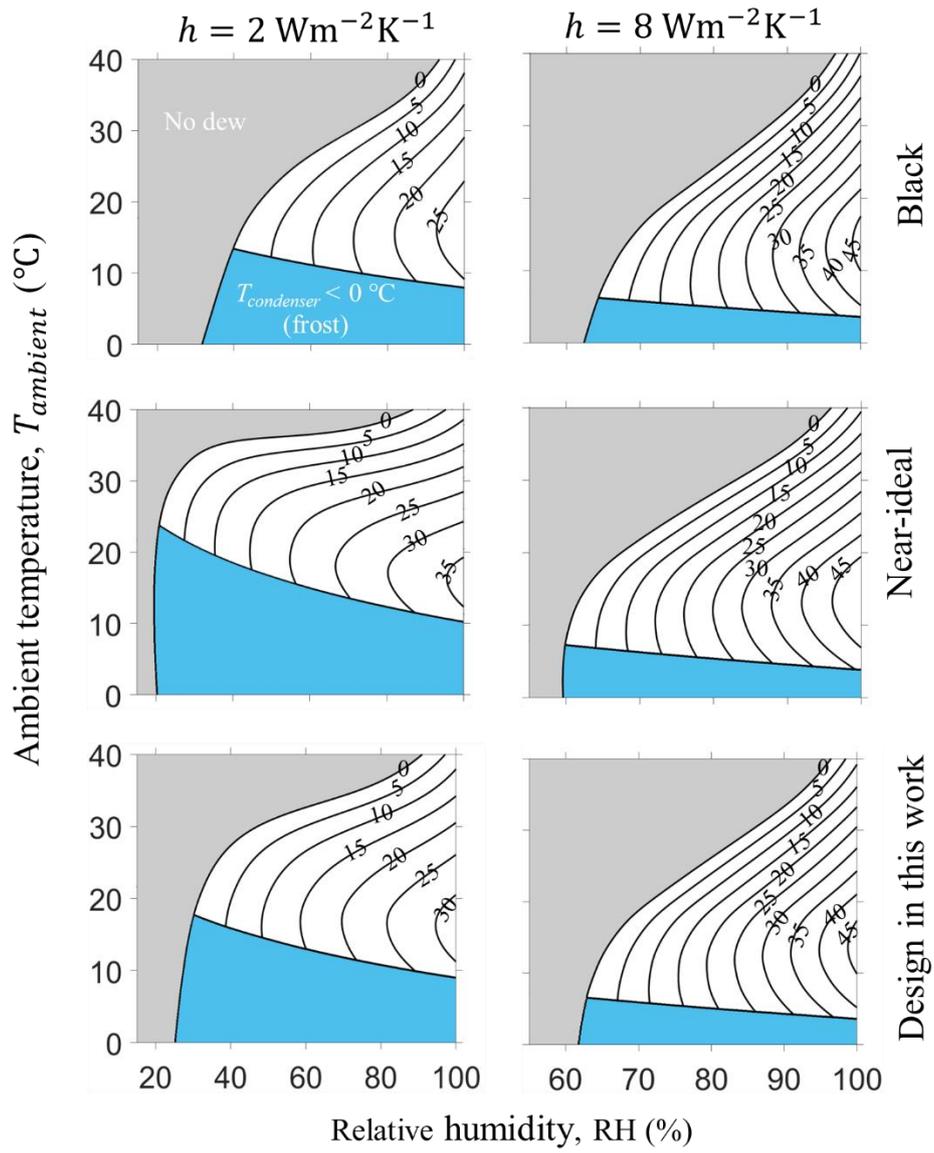

**Figure S1.** Contour plots of the condensation mass flux, $\dot{m}''$, as a function of RH (*x*-axis) and $T_{ambient}$ (*y*-axis) for the three condensers (rows) under the two representative scenarios (columns): $h = 2$ Wm$^{-2}$K$^{-1}$ and 8 Wm$^{-2}$K$^{-1}$.

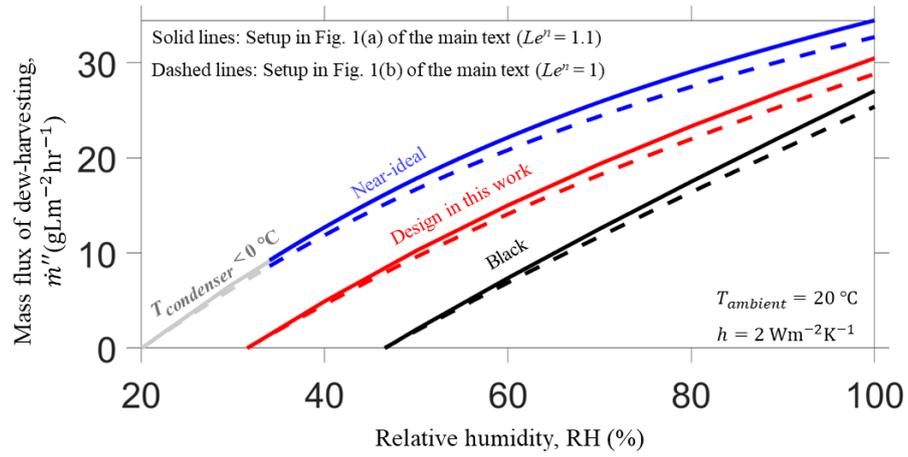

**Figure S2**. Comparison of the condensation mass flux, $\dot{m}''$ between the two setups in Fig. 1a and 1b of the main text. The difference is within 6% for all the three condensers. Grey lines represent $\dot{m}''$ values that would have been reached if dew formed. However, in reality frost forms instead, because under these conditions, $T_{condenser}$ is below 0℃.

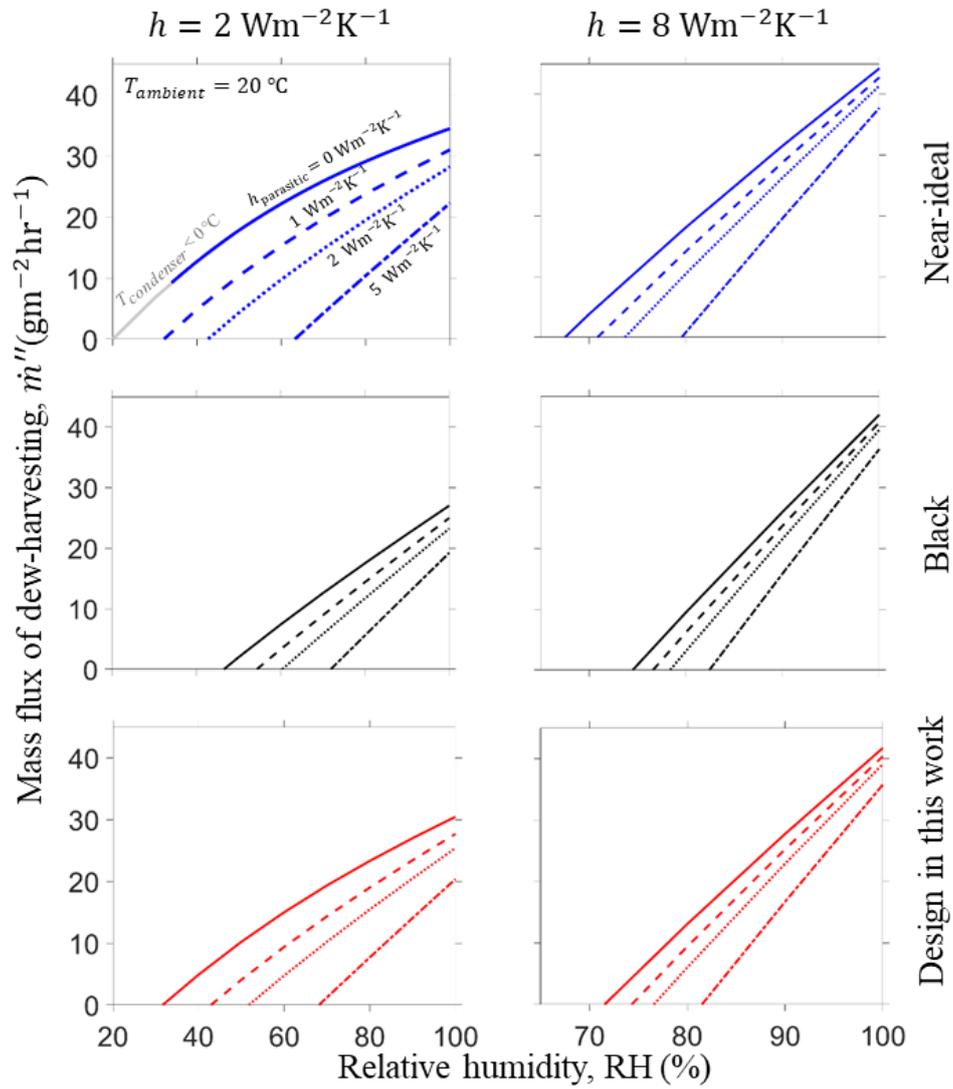

**Figure S3.** Effect of the parasitic heat transfer coefficient, $h_{parasitic}$, on the condensation mass flux, $\dot{m}''$. Here, the parasitic heat refers to the heat loss through the lower surface of the condenser (Fig. 1a of the main text), or through the upper surface of the condenser (Fig. 1b of the main text). This parasitic heat reduces the net cooling flux for condensation, and thus reducing $\dot{m}''$, regardless of the type of the condenser and the value of the convection coefficient, $h$. Grey lines represent $\dot{m}''$ values that would have been reached if dew formed. However, in reality frost forms instead, because under these conditions, $T_{condenser}$ is below 0 °C.

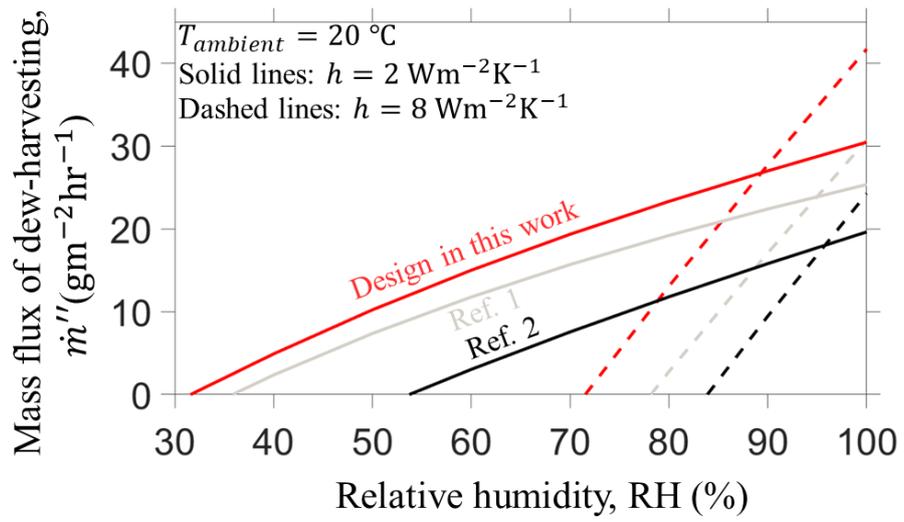

**Figure S4**. Comparison with previous designs [1-2] on the condensation mass flux, $\dot{m}''$. Design in this work (red), aiming at maximizing is apparently better than our previous designs (grey and black), which is optimized for radiative cooling only.

**Supporting Information Note 1:** Relation between the heat and mass transfer coefficients: derivation of Eq. (6) of the main text

We employ the strong analogy between the heat and mass transfer,[3-4] and analyze three scenarios.

First, we recall the textbook example of the forced external convection over a horizontal plate, which has a Nusselt number[4]

$$Nu = \frac{hL}{k} = 0.664(Re_L)^{1/2}(Pr)^{1/3}, \tag{S1}$$

in which $h$ and $k$ are the convection coefficient and the thermal conductivity of the air, $L$ is the characteristic length of the plate, and $Re$ and $Pr$ are the Reynold number and Prandtl number, respectively.

Correspondingly, the Nusselt number of the mass transfer is[4]

$$Nu_m = \frac{g_{m,H_2O} L}{\rho_{air} D_{H_2O,air}} = 0.664(Re_L)^{1/2}(Sc)^{1/3}, \tag{S2}$$

which just simply replaces the Prnadtl number, $Pr$, with the Schmidt number, $Sc$. $\rho_{air}$ is the density of air, and $D_{H_2O,air}$ is the mass diffusion coefficient of the water vapor in the air.

Combining Eqs. (S1-S2), we obtain

$$\frac{Nu_m}{Nu} = \left(\frac{\frac{g_{m,H_2O} L}{\rho_{air} D_{H_2O,air}}}{\frac{hL}{k}}\right) = \left(\frac{Sc}{Pr}\right)^{\frac{1}{3}}, \tag{S3}$$

in which the first bracket gives $c_p \frac{g_{m,H_2O}}{h} Le$, and the second bracket gives $(Le)^{\frac{1}{3}}$. Therefore, we arrive at

$$g_{m,H_2O} = \frac{h}{c_p}(Le)^{-\frac{2}{3}}, \tag{S4}$$

which gives the relation between the mass transfer coefficient, $g_{m,H_2O}$, and the heat transfer coefficient, $h$, for the forced external convection over a horizontal plate (2nd column of Table 2 of the main text).

The relation between $g_{m,H_2O}$, and $h$ for an inclined plate is exactly the same with Eq. S4, since the only difference in $Nu$ and $Nu_m$ is the replacement of the constant 0.664 with different constants according to the inclined angles.[5]

Second, we derive a similar relation for the natural convection over a horizontal or an inclined plate. One commonly used correlation is[3]

$$Nu = \frac{hL}{k} = 0.27 Ra_{L_c}^{1/4}, \tag{S5}$$

in which $L_c$ is the modified characteristic length of the plate, and the Rayleigh number is

$$Ra_{L_c} = \frac{g\cos\theta L_c^3}{\nu\alpha} \frac{\Delta T}{T}, \tag{S6}$$

where $\theta$ is the angle measured from the vertical direction to the direction along the inclined plate.

The corresponding Nusselt number of the mass transfer is

$$Nu_m = \frac{g_{m,H_2O} L}{\rho_{air} D_{H_2O,air}} = 0.27 Ra_{m,L_c}^{1/4}, \tag{S7}$$

which just simply replaces the heat transfer Rayleigh number, $Ra_{L_c}$, with the mass transfer Rayleigh number,[4]

$$Ra_{m,L_c} = \frac{g\cos\theta L_c^3}{\nu D_{H_2O,air}} \frac{\Delta \rho_{air}}{\rho_{air}}. \tag{S8}$$

Combining Eqs. (S5-S8), we obtain

$$\frac{Nu_m}{Nu} = \left(\frac{\frac{g_{m,H_2O} L}{\rho_{air} D_{H_2O,air}}}{\frac{hL}{k}}\right) = \left(\frac{Ra_{m,L_c}}{Ra_{L_c}}\right)^{\frac{1}{4}}, \tag{S9}$$

in which the first bracket again gives $c_p \frac{g_{m,H_2O}}{h} Le$, and the second bracket gives $(Le)^{\frac{1}{4}}$, which utilizes the Ideal Gas Law. Therefore, we arrive at

$$g_{m,H_2O} = \frac{h}{c_p}(Le)^{-\frac{3}{4}}, \tag{S10}$$

which gives the relation between $g_{m,H_2O}$ and $h$ for the natural convection over a horizontal or an inclined plate (1st column of Table 2 of the main text).

In the last scenario, we derive the relation for the internal flow through pipes with various shapes of the cross section. In this case, the heat and mass transfer shares with a same constant Nusselt number

$$Nu = Nu_m = \text{constant}, \tag{S11}$$

where the constant depends on the shape of the cross section.

Therefore, we obtain

$$\frac{Nu_m}{Nu} = \left(\frac{\frac{g_{m,H_2O}L}{\rho_{air}D_{H_2O,air}}}{\frac{hL}{k}}\right) = 1, \tag{S12}$$

in which the first bracket again gives $c_p \frac{g_{m,H_2O}}{h} Le$. Thus, we arrive at

$$g_{m,H_2O} = \frac{h}{c_p}(Le)^{-1}, \tag{S13}$$

which gives the relation between $g_{m,H_2O}$ and $h$ for the forced internal convection through a pipe with various shapes of the cross section (3rd column of Table 2 of the main text).

In Summary, Eqs. S4, S10, and S13 give the relation between $g_{m,H_2O}$ and $h$ for three different scenarios, which are characterized by different $n$ values, as outlined in Table 2 of the main text. We note that the differences among these scenarios are less than 6%, as shown in Fig. S2 for a comparison between the setups of Fig. 1(a) (2nd column of Table 2) and Fig. 1(b) (3rd column of Table 2) of the main text.

**Supporting Information Note 2:** Effect of the parasitic heat transfer coefficient

In Eq. (1) of the main text, we set the parasitic heat transfer coefficient, $h_{parasitic}$ = 0, to reach the fundamental limits. In this note, we analyze the effect of a nonzero $h_{parasitic}$. As shown in Fig. S3, the mass flux of the dew-harvesting, $\dot{m}''$, decreases as the increase of $h_{parasitic}$ for any condenser with any convection coefficient, $h$, which is consistent with our insight in the main text.

**Supporting Information Note 3:** Comparison with other photonic designs

The photonic design in this work represents the first design aiming at maximizing the mass flux of the dew-harvesting, $\dot{m}''$, in contrast with those designs merely for radiative cooling, e.g. in Refs. 1 and 2. As expected, the $\dot{m}''$ of this design surpasses our previous designs, as shown in Fig. S4.